\documentclass[showpacs,amssymb,amsmath,nobibnotes, aps,prl,secnumarabic
,twocolumn%
]{revtex4}
\pdfoutput=1
\usepackage{bm}
\usepackage{natbib}
\usepackage{graphicx}

\begin{document}
\textfloatsep 10pt

\title{Turbulence damping as a measure of the flow dimensionality}

\author{M. Shats}
\email{Michael.Shats@anu.edu.au}
\author{D. Byrne}
\author{H. Xia}

\affiliation{Research School of
Physics and Engineering, The Australian National
University, Canberra ACT 0200, Australia}

\date{\today}

\begin{abstract}
The dimensionality of turbulence in fluid layers determines their properties.
We study electromagnetically driven flows in finite depth fluid layers and show that
eddy viscosity, which appears as a result of three-dimensional motions, leads to increased bottom damping.
The anomaly coefficient, which characterizes the deviation of damping from the one derived using a quasi-two-dimensional model, can be used as a measure of the flow dimensionality. Experiments in turbulent layers show that when the anomaly coefficient becomes high, the turbulent inverse energy cascade is suppressed. In the opposite limit turbulence can self-organize into a coherent flow.

\end{abstract}


\maketitle

Fluid layers represent a broad class of flows whose depths are much smaller than their horizontal extents, for example, planetary atmospheres and oceans. A discovery of the upscale energy transfer in two-dimensional (2D) turbulence \cite{Kraichnan1967} gave new insight into the energy balance in turbulent layers. The inverse cascade transfers energy from smaller to larger scales thus allowing for turbulence self-organization. This is in contrast with three-dimensional (3D) turbulence where energy is nonlinearly transferred towards small scales (direct cascade).

Real physical layers differ from the ideal 2D model since they have finite depths and non-zero dissipation. The effect of the layer thickness on turbulence driven by 2D forcing has been studied in 3D numerical simulations \cite{Smith_PRL_1996,Celani_PRL_2010}. It has been shown that in ``turbulence in more than two and less than three dimensions", the injected energy flux splits into the direct and inverse parts. At ratios of the layer depth $h$ over the forcing scale $l_f$ above $h/l_f \sim 0.5$ the inverse energy cascade is greatly reduced. When the inverse energy flux is suppressed, the energy injected into the flow is transferred towards small scales by the direct cascade, developing the Kolmogorov $k^{-5/3}$ spectrum at $k>k_f$. This result illustrates that 2D and 3D turbulence may coexist.

2D/3D effects have been studied in electromagnetically driven flows using two main schemes to force the fluid motion. In liquid metals placed in the vertical homogeneous magnetic field the flow is forced by applying spatially varying electric field which generates $J\times B$ forces. In such magnetohydrodynamic (MHD) flows 2D properties are enforced by the magnetic field and the 3D behavior is restricted to a very thin \textit{Hartmann layer} \cite{Sommeria1986}. The deviations from 2D in such flows may be due to the finite resistivity in very thick layers \cite{Thess2007,Klein2010}. Another class of experiments employs spatially periodic magnetic field crossed with the constant horizontal electric current to produce interacting vortices \cite{Paret1998Intermittency, Boffetta_EPL_2005,Xia2009}. In this case the thickness of the Hartmann layer exceeds the layer depth and 2D/3D effects are determined by the factors which are different from those in MHD flows, for example, by a density stratification.

The 3D effects are closely related to the energy dissipation in the layers. This connection however is not fully understood in experiments. The measured flow damping rates are often compared with those derived from a quasi-2D model \cite{Dolzhanskii_JFM_1992,Hansen_PRE_1998} which assumes no vertical motions within the layer. In thin layers, the agreement is usually within a factor of 2 \cite{Boffetta_EPL_2005, Danilov_PRE_2002}. However in some experiments a much better agreement with the quasi-2D model was observed \cite{Clercx_van_Heijst_2003}. This contradicts recent claims about the intrinsic three-dimensionality of the flows in thin layers of electrolytes \cite{Akkermans2008, AkkermansPRE2010}. There is a need to clarify this.

Physical three-dimensionality of the flow is determined by the amount of 3D motion in the layer. This motion may naturally develop in the layer, as in \cite{Celani_PRL_2010}, but it can also be injected into the flow by non-2D forcing or it can be generated by the shear-driven instabilities in the boundary layer. In this case, the critical layer thickness cannot be used as a practical criterion of the 2D/3D transition since it will vary depending on the source of 3D motion. The transition from 2D to 3D, which marks a fundamental change in the energy transfer, needs to be characterized quantitatively, in other words, it is necessary to find a measure of the flow dimensionality which would help to predict turbulence behavior.

In this Letter we show that eddy viscosity increases damping in finite-depth fluid layers compared with the quasi-2D model prediction. This increase can be used as the measure of the flow dimensionality which allows to evaluate the likelihood of the inverse energy cascade and of turbulence self-organization. We also show that the increased degree of three-dimensionality leads to the suppression of the turbulent cascades.

In these experiments turbulence is generated via the interaction of a large number of electromagnetically driven vortices \cite{ShatsPRE2005, Xia2008, Xia2009}. The electric current flowing through a conducting fluid layer interacts with the spatially variable vertical magnetic field produced by arrays of magnets placed under the bottom. In this paper we use a $30 \times 30$ array of magnetic dipoles (8 mm apart) for the turbulence studies requiring large statistics. For the studies of vertical motions, a $6 \times 6$ array of larger magnets (25 mm separation) is used. The flow is visualized using seeding particles, which are suspended in the fluid, illuminated using a horizontal laser slab and filmed from above. Particle image velocimetry (PIV) is used to derive turbulent velocity fields. The flow is generated either in a single layer of electrolyte ($Na_2SO_4$ water solution), or in two immiscible layers of fluids (electrically neutral heavier liquid at the bottom, electrolyte on top). Shortly after the current is switched on, $J \times B$ driven vortices interact with each other forming complex turbulent motion characterized by a broad wave number spectrum. The steady state is reached within tens of seconds.

\begin{figure}
\includegraphics[width=5.0 cm]{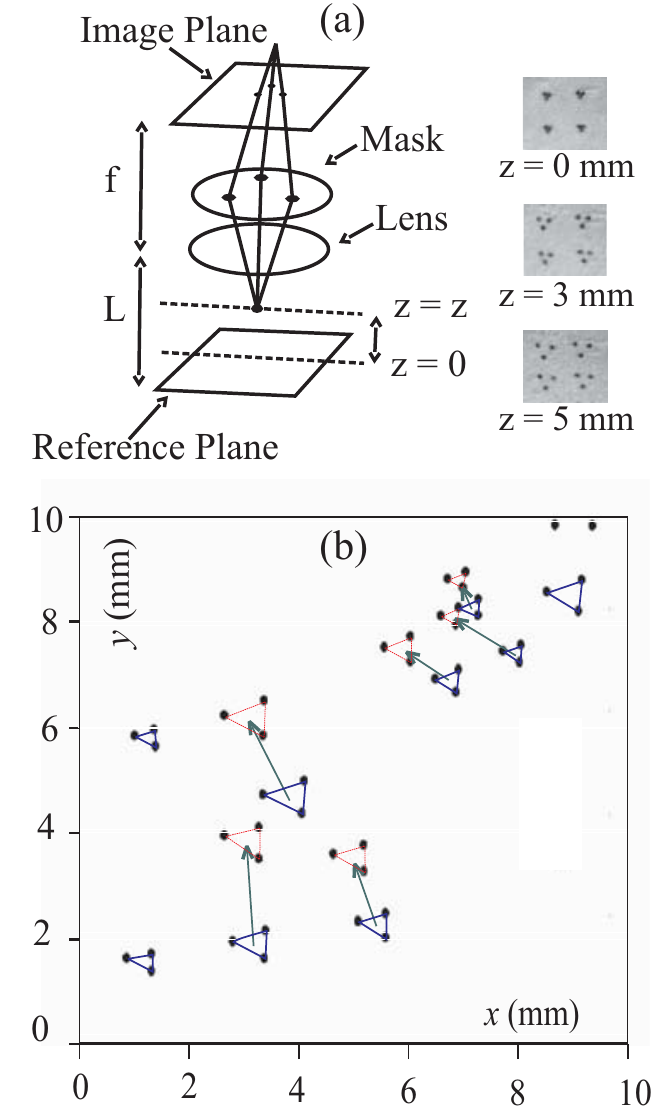}
\caption{\label{DefocPIV} Schematic of the defocusing particle image velocimetry technique.}
\end{figure}

To study vertical motions in single electrolyte layers, vertical laser slabs are used to illuminate the flow in the $y-z$ plane. Streaks of the seeding particles within the slab are filmed with the exposure time of 1 s. Quantitative measurements of the horizontal and vertical velocities are performed using defocusing PIV technique. This technique, was first described in \cite{Willert1991}, but had never been used in turbulence studies. It allows measurements of 3D velocity components of seeding particles using a single video camera with a multiple pinhole mask (three pinholes constituting a triangle are used here). A schematic of the method is shown in Fig.~\ref{DefocPIV}. An image of a particle placed in the reference plane at $z = $0 (where the particle is in focus) corresponds to a single dot in the image plane. As the particle moves vertically away from the reference plane, the light passes through each pinhole in the mask and reaches three different positions on the image plane. The distances between the triangle vertices in the image plane are used to decode $z$-positions of the particles. The $xy$-components of velocity are determined using a PIV/PTV hybrid algorithm to match particle pairs from frame to frame. This process is illustrated in Fig.~\ref{DefocPIV}. The technique allows to resolve vertical velocities above $<V_z>_{RMS} \ge 0.5$ mm/s. The imaged area in this experiment is $5 \times 5$ cm$^2$. On average about 50 particles (triangles) are tracked in two consecutive frames. Derived velocities are then averaged over about 100 of the frame pairs to generate converged statistics of the mean-square-root velocities $<V_{x,y,z}>_{RMS}$.

\begin{figure}
\includegraphics[width=8.0 cm]{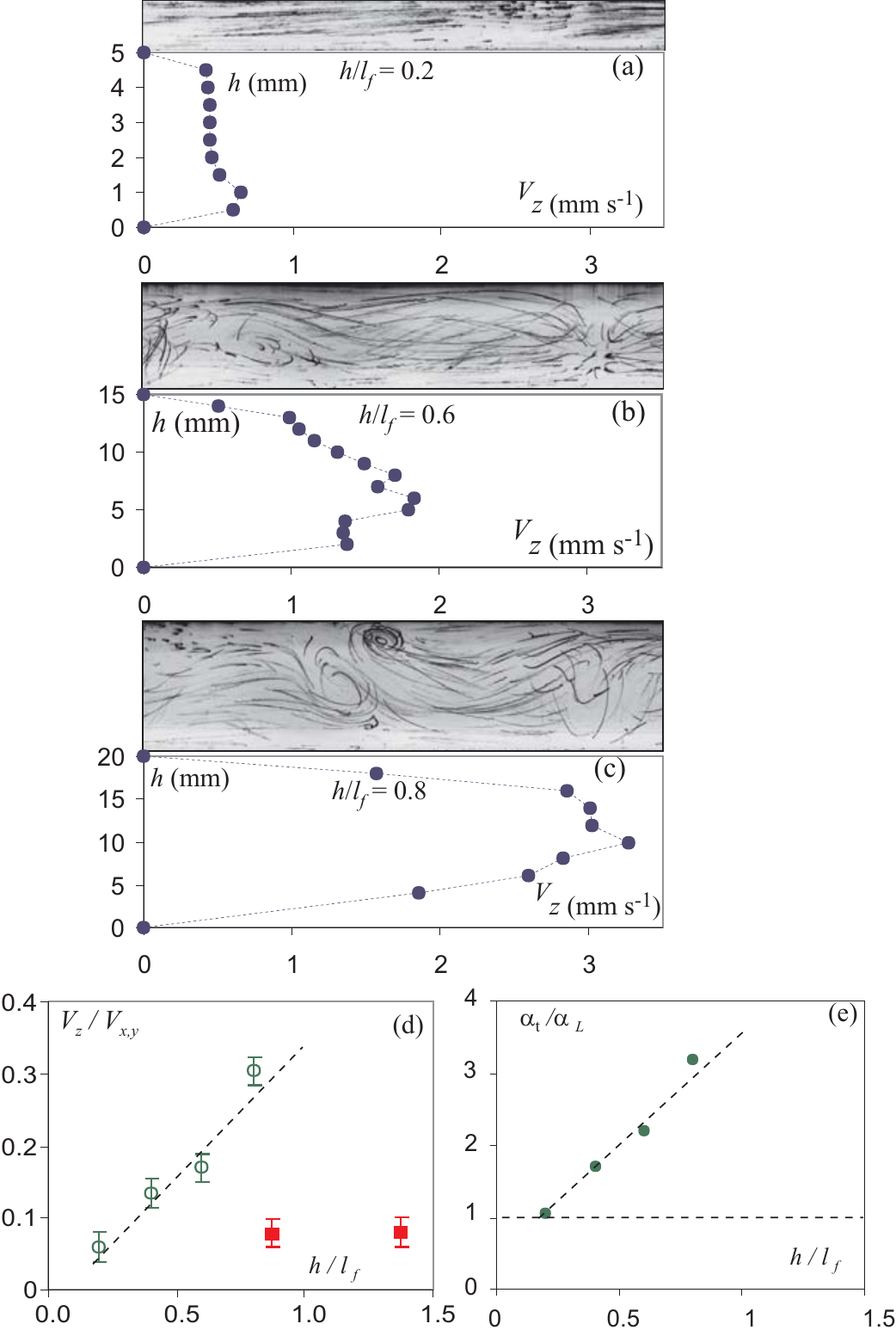}
\caption{\label{Vertical_motion} Particle streaks filmed with an exposure time of 1 s (top panels) and the distribution of the vertical velocity fluctuations (rms) over the layer thickness (bottom panels) in single layers: (a) $h= 5$ mm; (b) $h= 15$ mm; (c) $h= 20$ mm. (d) Ratio of rms vertical to the rms horizontal velocity as a function of the normalized layer thickness $h/l_f$ in a single (open circles) and in a double (solid squares) layer configurations. (e) $\alpha_t/\alpha_L$ versus $h/l_f$.}
\end{figure}

Figures~\ref{Vertical_motion}(a-c) show particle streaks and corresponding vertical velocity profiles $V_z(z)$ for different layer depths. To keep forcing approximately constant, the electric current is increased proportionally to the layer thickness (constant current density). To obtain better vertical spatial resolution, a $6 \times 6$ array of larger magnets is used. For the layers thicknesses of up to 30 mm, a range of $h/l_f = 0.2 - 1.2$ is achieved. Particle streaks show reasonably 2D motion in a thin (5 mm) layer, Fig.~\ref{Vertical_motion}(a). Vertical velocity is small over most of the layer thickness and is close to the resolution of the technique, $<V_z>_{RMS} \sim 0.5$ mm/s. As the layer thickness is increased, 3D motions develop. The corresponding vertical velocities increase up to $\sim 4$ mm/s, Figs.~\ref{Vertical_motion}(b,c). Fig.~\ref{Vertical_motion}(d) shows the ratio of vertical to horizontal velocities as a function of the normalized layer thickness. In single layers this ratio increases approximately linearly with $h/l_f$ reaching over $<V_z>/<V_{x,y}>$ $ = 0.3$ at $h/l_f = 0.8$. In stratified double layers this ratio is substantially smaller, $<V_z>/<V_{x,y}> $ $\le 0.08$ (solid squares in Fig.~\ref{Vertical_motion}(d)), suggesting that the flow in a double layer configuration is much closer to 2D.

In the absence of 3D motions, the flow in the layer is damped due to molecular viscosity. A decay of horizontal velocity $V_{x,y}(z,t)$ in the quasi-2D flow due to the bottom friction is described by the diffusive type equation $\partial V_{x,y}/ \partial t = \nu \partial^2 V_{x,y} / \partial z^2$, which together with the boundary conditions $V_{x,y}(z=0,t)=0$ and $\partial V_{x,y}(z=h,t)/\partial z=0$ gives the characteristic inverse time of the energy decay, e.g. \cite{Dolzhanskii_JFM_1992}:
\begin{equation}
\label{alpha_L}
\alpha_L = \nu \pi^2 / 2h^2.
\end{equation}

\noindent Here $\nu$ is the kinematic viscosity.

The onset of 3D turbulent eddies in thicker layers should lead to a vertical flux of horizontal momentum and faster dissipation of the flow.
Such a flux is related to the mean vertical velocity gradient $\partial V_{x,y}/ \partial z$ \cite{Meteorology_textbook}:
\begin{equation}
\label{Mom_flux}
<\tilde{V}_{x,y}\tilde{V}_{z}>=-K \frac{\partial V_{x,y}}{\partial z}.
\end{equation}

\noindent Here $K$ is the eddy (turbulent) viscosity coefficient. By assuming that fluctuations of vertical and horizontal velocities are well correlated, we can estimate the eddy viscosity coefficient using the defocusing PIV data as $K \approx <\tilde{V}_{x,y}><\tilde{V}_{z}>(\partial V_{x,y}/ \partial z)^{-1}$. Then the damping rate can be estimated using the contribution of both molecular and the eddy viscosities, $\alpha_t= (\nu+K) \pi^2 / 2h^2$. The ratio of thus calculated damping rate to the linear damping $\alpha_L$ (\ref{alpha_L}) is shown in Fig.~\ref{Vertical_motion}(e).

The damping should become anomalous ($\alpha_t / \alpha_L >1$) above some critical layer thickness of $h / l_f \approx 0.3$.
According to Fig.~\ref{Vertical_motion}(e) this anomaly should increase linearly with the increase in $h/l_f$.

\begin{figure}
\includegraphics[width=4.0 cm]{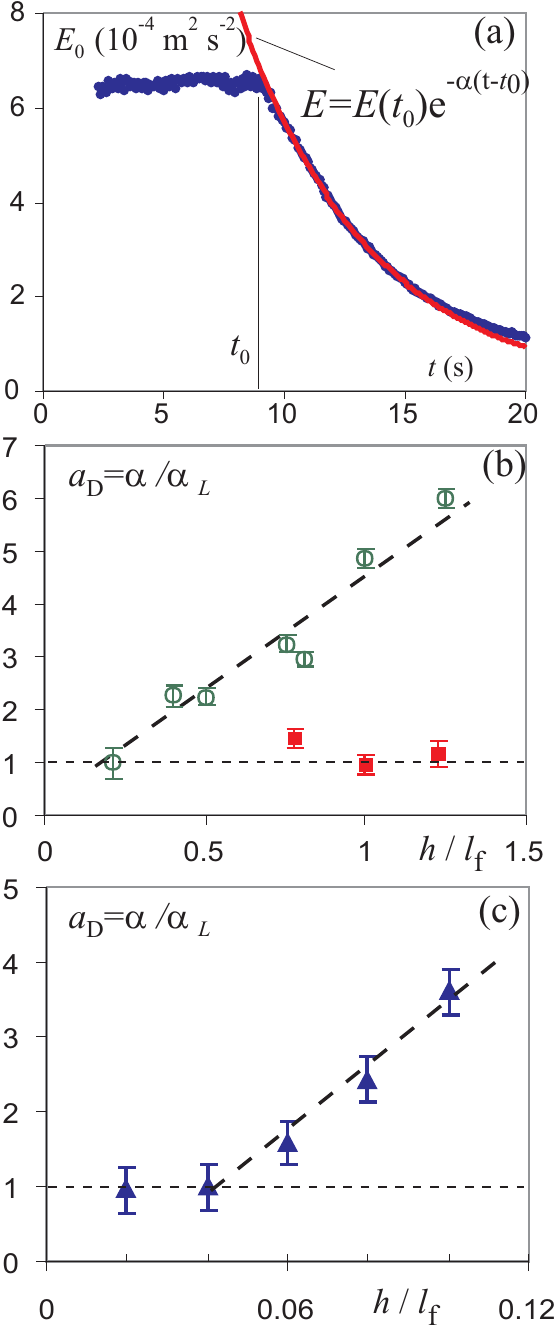}
\caption{\label{Damping_measurement} (a) Decay of the flow energy in a single layer, $h= 10$ mm; (b) Energy damping rate normalized by the viscous quasi-2D damping rate $a_D=\alpha/\alpha_L$, as a function of $h/l_f$. Open circles refer to single layers, solid squares were obtained in the double layer configurations. (c) The damping anomaly coefficient $a_D$ versus $h/l_f$ for the case of a strong large-scale vortex (100 mm diameter, $V_{x,y}^{max}=16$ mm/s).}
\end{figure}

Direct measurements of damping were performed to test that eddy viscosity increases the dissipation above its quasi-2D value (\ref{alpha_L}) in layers thicker than $h/l_f > 0.2$. The flow is forced by a $30 \times 30$ magnet array . The bottom drag is derived from the energy decay of the steady flow. After forcing is switched off, the mean flow energy exponentially decays in time with a characteristic time constant $\alpha$, as shown in Fig.~\ref{Damping_measurement}(a). We compare the energy damping rate measured in a single layer of different depths with the linear damping rate. Fig.~\ref{Damping_measurement}(b) shows the anomaly coefficient $a_D=\alpha / \alpha_L$ as a function of the normalized layer thickness $h/l_f$. In the thinnest layer ($h \approx 1.7$ mm, $h/l_f \approx 0.21$) the damping rate coincides with the linear damping rate (\ref{alpha_L}). However for thicker layers the damping anomaly is higher, such that $a_D$ increases linearly with $h$ reaching $a_D=6$ at $h/l_f = 1.25$.

Measurements of the damping show that the anomaly coefficient $a_D$ in Fig.~\ref{Damping_measurement}(b) agrees very well with the anomaly estimated using the eddy viscosity derived from (\ref{Mom_flux}), Fig.~\ref{Vertical_motion}(e).
In the double layer experiments however, $a_D$ is substantially lower, as shown by the solid squares in Fig.~\ref{Damping_measurement}(b). This is not surprising in the light of the result of Fig.~\ref{Vertical_motion}(d) (solid squares) which shows substantially less 3D motion in double layers.

The above results are related to low forcing levels, when 3D eddies are generated due to the finite layer thickness, as in \cite{Celani_PRL_2010}.
However, electromagnetic forcing, which is maximum near the bottom in the single layer experiments (magnets underneath the fluid cell), may inject 3D eddies into the flow from the bottom boundary layer at higher forcing levels. Figure~\ref{Damping_measurement}(c) shows the damping anomaly coefficient $a_D$ measured in the flow driven by a single strong large magnetic dipole. A single large-scale vortex is produced, whose diameter is about 100 mm and the maximum horizontal velocity is about 16 mm/s. As the layer thickness is increased from 2 to 10 mm ($h/l_f=0.02-0.1$) while keeping the current density constant, the anomaly coefficient increases up to $a_D = 3.6$ due to the increase in the vertical velocity fluctuations. Thus, turbulent bottom drag may occur in relatively thin layers at stronger forcing.

\begin{figure}
\includegraphics[width=4.5 cm]{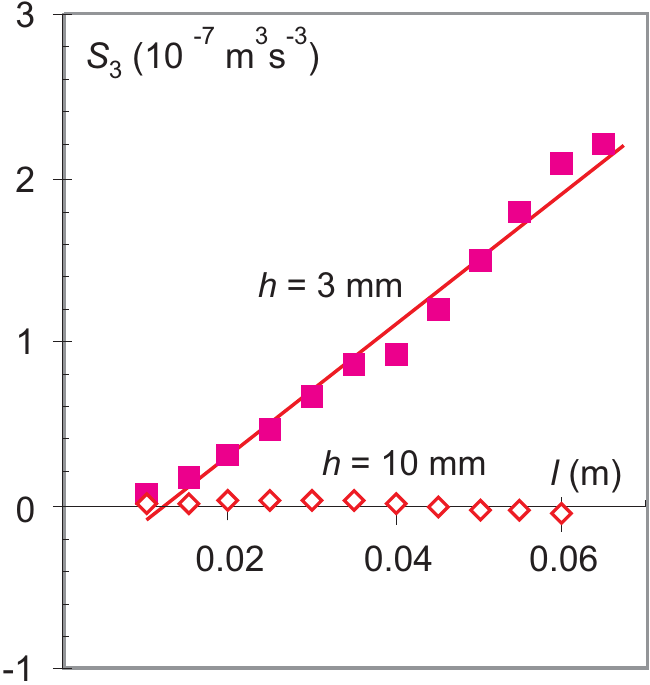}
\caption{\label{Inverse_cascade} Third-order structure functions measured in a thin layer, $h =$ 3 mm (solid squares), and in a thick layer $h =$ 10 mm (open diamonds). The forcing scale $l_f \approx 8$ mm.}
\end{figure}

Now we test if the increased three-dimensionality, as characterized by $a_D$, leads to the suppression of the inverse energy cascade.
The inverse energy cascade can be detected by measuring the third-order structure function $S_3$ and by using the Kolmogorov flux relation which predicts linear dependence of $S_3$ on the separation distance $l$, $S_3=\epsilon l$. Here $\epsilon$ is the energy flux in $k$-space. It has been shown that in thin stratified layers $S_3$ is positive and it is a linear function of $l$, as expected for 2D turbulence \cite{Xia2009}. Figure~\ref{Inverse_cascade} shows third-order structure functions measured in a single layer of electrolyte for two layer depths, $h=$ 3 and 10 mm. In the 3 mm layer, $S_3$ is a positive linear function of $l$, while in the 10 mm layer $S_3$ is much smaller, indicating very low energy flux in the inverse energy cascade. The damping anomaly in the 3 mm layer is $a_D \approx 2$, while for the 10 mm layer it is high, $a_D \approx 5$. Since in this experiment, the forcing is 2D and it is relatively weak (no secondary instabilities in the boundary layer), this result is in agreement with numerical simulations \cite{Celani_PRL_2010} which show strong suppression of the inverse energy cascade above $h/l_f \ge 0.5$. The 3 mm layer corresponds to $h/l_f \approx 0.38$, while for the 10 mm layer $h/l_f \approx 1.25$. We do not observe however any signatures of the direct energy cascade range, $E_k \propto k^{-5/3}$ at $k>k_f$ in the 10 mm layer. Instead, the spectrum is much steeper than the usual $k^{-3}$ enstrophy range. This is probably due to the fact that the Reynolds number in this experiment is not sufficient to sustain 3D direct turbulent cascade.

Summarizing, we demonstrate for the first time that increased three-dimensionality of flows in layers can be characterized by the anomalous damping coefficient $a_D$. We show that the increase in $a_D$ correlates with the suppression of the inverse energy cascade. On the other hand, a strong reduction in $a_D$, which can be achieved in the double layer configuration, correlates well with the observation of the inverse energy cascade and spectral condensation of turbulence into a flow coherent over the entire domain \cite{Paret1998Intermittency,ShatsPRE2005,Xia2009}.

\begin{acknowledgments}
The authors are grateful to H. Punzmann and V. Steinberg for useful discussions. This work was supported by the Australian Research Council's Discovery Projects funding scheme (DP0881544).
\end{acknowledgments}

\end{document}